\documentclass[prb,preprint]{revtex4-1}
\usepackage{graphicx}
\usepackage{amsmath,amssymb,amsfonts}
\usepackage{natbib}
\usepackage{color}

\newcommand{\be}{\begin{equation}}
\newcommand{\ba}{\begin{eqnarray}}
\newcommand{\ee}{\end{equation}}
\newcommand{\ea}{\end{eqnarray}}

\begin{document}

\title{How does a magnetic trap work?}

\author{J. P\'erez-R\'{\i}os}
\email[Current address: Department of Physics, Purdue University,
525 Northwestern Avenue, West Lafayette, IN 47907-2036, USA.]{\
E-mail: jperezri@purdue.edu} \affiliation{Laboratoire Aim\'{e}
Cotton, CNRS, Universit\'{e} Paris-Sud, B\^{a}t 505, 91404 Orsay,
France}

\author{A. S. Sanz}
\affiliation{Instituto de F\'{\i}sica Fundamental (IFF--CSIC),
Serrano 123, 28006 Madrid, Spain}
\affiliation{Department of Physics and Astronomy, University College
London, Gower Street, London WC1E 6BT, United Kingdom}

\date{\today}

\begin{abstract}
Magnetic trapping is a cornerstone for modern ultracold physics and
its applications (e.g., quantum information processing, quantum
metrology, quantum optics, or high-resolution spectroscopies). Here
a comprehensive analysis and discussion of the basic physics behind
the most commonly used magnetic traps in Bose-Einstein condensation
is presented.
This analysis includes the
quadrupole trap, the time-averaged orbiting potential trap, and the
Ioffe-Pritchard trap. It is shown how the trapping conditions and
efficiency of these devices can be determined from simple derivations
based on classical electromagnetism, even though they operate on
quantum objects.
\end{abstract}

%\pacs{05.10.Gg, 05.40.-a, 68.43.-h}

%02.50.Ey Stochastic processes
%05.10.Gg Stochastic analysis methods (Fokker-Planck, Langevin, etc.)
%05.40.-a Fluctuation phenomena, random processes, noise, and Brownian motion
%68.43.-h Chemisorption/physisorption: adsorbates on surfaces
%82.20.Db Transition state theory and statistical theories of rate constants

\maketitle

%%%%%%%%%%%%%%%%%%%%%%%%%%%%%%%%%%%%%%%%%%%%%%%%%%%%%%%%%%%%%%%%%%%%%%%
%%%%%%%%%%%%%%%%%%%%%%%%%%%%%%%%%%%%%%%%%%%%%%%%%%%%%%%%%%%%%%%%%%%%%%%

\section{Introduction}
\label{sec1}

In 1995 the first Bose-Einstein condensates (BECs) were synthesized
experimentally.\cite{Anderson-95,Davis-95,Bradley-95} Ever since,
ultracold physics as well as its applications (e.g., quantum
information processing, quantum metrology, quantum computation,
quantum optics, high-resolution spectroscopies, etc.) have undergone
a remarkable development.\cite{Carr-2009} In the ultracold regime,
the particles (atoms or molecules) that constitute the condensate
have to behave as a whole, i.e., they must constitute a {\it
coherent} cloud, as if they all just formed a single particle.
In order to observe this behavior, the thermal de Broglie wavelength,
$\lambda_{\rm dB}$, of the species considered has to be at least of
the order of magnitude of the cloud size. This means that
$\lambda_{\rm dB} \gtrsim n^{-1/3}$, with $n$ being the particle
density.
Under these circumstances, the condensate dynamics
will be completely dominated by quantum
effects.\cite{Griffin,Dalfovo-99,Legget-2001,Pethick}

One of the methods utilized to reach the ultracold regime is the
process known as {\it evaporative cooling}.\cite{Hess-86,Ketterle-96}
According to this method, first an ensemble of particles (cloud) that
follows a Maxwell-Boltzmann velocity distribution is confined within
a {\it magnetic trap} (it could also be a {\it dipolar trap}).
Then, the most energetic particles
will escape from the trap due to their higher energy, while the less
energetic ones
will remain inside it. The net effect is thus an effective cooling,
similar to the process leading coffee in a cup to cool down, for
example. In order to reach thermalization by inter-particle elastic
collisions at very low temperatures, the walls of the trap are also
lowered. This produces an increment of the particle loss rate, which
eventually leads the remaining particles to condensate.

Magnetic trapping has its origins in plasma physics, where the
confinement of charged particles\cite{Fusion,Goldstone} is one of
the main steps in the control of fusion reactions that take place
inside plasmas.
The first technique employed to trap and confine charged particles
within this field was the so-called magnetic bottle
configuration, which results from the combination of two magnetic
mirrors. In this configuration the trapped particles bounce back and
forth periodically between the two mirrors (turning points), where
the intensity of the magnetic field is relatively high, thus
preventing the particles from escaping.\cite{Allen-62}

The magnetic bottle configuration described above cannot be
directly applied to BECs, because they are neutral particles.
Nevertheless, magnetic trapping in this case still relies on the same
basic idea, although it involves a different trapping mechanism, namely
the combination of the Zeeman effect with Earnshaw's
theorem.\cite{Earnshaw,Wing-84} On the one hand, the
Zeeman effect originates the splitting of the particle energy levels
in the presence of a magnetic field due to the interaction between
this field and the particle magnetic dipole moment. On the other
hand, Earnshaw's theorem determines the properties of a particular
magnetic field configuration, thus playing a key role in the
understanding of neutral particle trapping.

With this in mind, it is then clear that a proper knowledge of the
functioning of neutral-particle magnetic trapping is fundamental to
understanding modern ultracold physics.\cite{Wieman-96,Carr-2009}
The motivation and purpose of this work is to explain the essential
elements of magnetic trapping in simple terms.
In particular, we have focused on basic physical properties of the
three most relevant configurations used in this research field, namely
the quadrupole trap, the time-averaged orbiting potential (TOP) trap,
and the Ioffe-Pritchard trap.
As it is shown, the trapping conditions and efficiency of
these devices can be determined from simple derivations obtained
from classical electromagnetism. From a pedagogical viewpoint, we
consider that this has a potential interest in elementary courses on
classical electromagnetism, for it relates in an easy fashion this
field of physics with quantum mechanics through the quantum nature of
BECs (actually, without making a explicit use of quantum mechanics).

The work is organized as follows. To be self-contained,
in Sec.~\ref{sec2} we present a brief account on the role of the
Zeeman effect and Earnshaw's theorem in magnetic trapping. The
description of the traps mentioned above and their
properties, such as performance and efficiency, are introduced in
Sec.~\ref{sec3}, while a comparative analysis is given in
Sec.~\ref{sec4}. Usually we find in the literature\cite{Pethick}
that the magnetic scalar field (or potential) for a trap is
introduced {\it ad hoc}, in terms of some plausibility arguments.
In this regard, our work complements a previous one by Gov {\it et
al.},\cite{gov}
also published in this journal, where the authors present a classical
and quantum-mechanical analysis of neutral-particle magnetic trapping.
The derivations presented in Sec.~\ref{sec3} are, however, very
simple and readily lead to the trapping condition. Finally, a series
of remarks drawn from this work are summarized in Sec.~\ref{sec5}.

%%%%%%%%%%%%%%%%%%%%%%%%%%%%%%%%%%%%%%%%%%%%%%%%%%%%%%%%%%%%%%%%%%%%%%%
%%%%%%%%%%%%%%%%%%%%%%%%%%%%%%%%%%%%%%%%%%%%%%%%%%%%%%%%%%%%%%%%%%%%%%%

\section{Zeeman effect and Earnshaw's theorem}
\label{sec2}

According to classical
electromagnetism,\cite{Jackson,Smythe,Jefimenko,Panofsky} the energy
of a magnetic dipole moment, ${\boldsymbol \mu}$, acted by an
external magnetic field, ${\bf B}$, is given by
\begin{equation}
 \label{eq-1}
 E = - \boldsymbol\mu \cdot {\bf B} .
\end{equation}
Quantum-mechanically this interaction produces a splitting of the
particle (quantum) energy levels, namely the so-called Zeeman
effect.\cite{Carrington} In atoms or molecules, the total magnetic
dipole moment is specified by the orbital angular momentum and spin
of the electrons from the valence shell. This quantity reads
as\cite{Carrington}
\begin{equation}
 \label{eq-3}
 \boldsymbol\mu = - g_s \mu_B {\bf S} - g_l \mu_B {\bf L} ,
\end{equation}
where $g_s \approx 2.002319$ is the electron spin
$g$-factor,\cite{odom:PRL:2006} ${\bf S}$ is the magnitude of the
electron spin angular momentum, $g_l=1$ is the electron orbital
$g$-factor, ${\bf L}$ is the magnitude of the electron orbital
angular momentum, and $\mu_B = e\hbar/2m_e = 0.67$~K/T is the Bohr
magneton.

An illustration of the Zeeman splitting in the case of a realistic
system, namely the $^{16}$O$_2(^3\Sigma_{g}^{-})$
molecule,\cite{Carrington} is displayed in Fig.~\ref{fig1}.
A detailed account of the calculations performed to obtain the results
displayed in this figure goes beyond the scope of the present work due
to the different technical (numerical) computational aspects involved.
Nonetheless, for reproducibility
purposes, the interested reader can find the relevant realistic
parameters used in these calculations in Mizushima's {\it The Theory of
Rotating Diatomic Molecules}.\cite{Mizushima}
These parameters are the rotational constant ($B_e = 1.438$~cm$^{-1}$),
the spin-rotation interaction constant ($\gamma_{\rm SR} = -0.0089$~cm$^{-1}$),
and the spin-spin coupling constant ($\gamma_{\rm SS} = 1.985$~cm$^{-1}$).

In this example, the splitting of the energy levels comes only from
the interaction between the molecular electronic spin and the
external magnetic field, since the valence-shell electrons for
molecules in a $^3\Sigma$ electronic state have a zero orbital
angular momentum.\cite{Carrington} As can be seen, two different
responses of the molecular internal states are found. There are
levels that increase their energy with the intensity of the magnetic
field (blue dashed lines), which are called {\it low-field seekers}
and are stable at low intensities (the concept of stability here
refers to the system energetic configuration, which should be
distinguished from the concept of dynamical stability; see below, in
Sec.~\ref{sec4}). On the
contrary, there are {\it high-field seeker} states (red solid
lines), which display an energy decrease as the field intensity
increases, thus being stable in regions of higher or more intense
magnetic fields.

%\begin{figure}
% \includegraphics[width=8.6cm]{J_PerezRios_Fig01.eps}
% \caption{\label{fig1}
%  Zeeman splitting of the $^{16}$O$_{2}(^{3}\Sigma_{g}^{-})$ molecular
%  energy levels (in temperature units) as a function of the magnetic
%  field intensity.
%  High-field and low-field seeker states are denoted, respectively,
%  by red solid lines and blue dashed lines; gray dotted lines
%  illustrate the appearance of Zeeman splitting in higher energy
%  levels.
%  These results have been obtained using realistic values in the
%  simulation:\cite{Mizushima}
%  $B_e = 1.438$~cm$^{-1}$ (rotational constant), $\gamma_{\rm SR} =
%  -0.0089$~cm$^{-1}$ (spin-rotation interaction), and $\gamma_{\rm SS}
%  = 1.985$~cm$^{-1}$ (spin-spin coupling).}
%\end{figure}

Taking these facts into account, a priori one would be willing to design
magnetic field configurations either with a local minimum to trap
low-field seeker states, or with a local maximum to trap high-field
seeker states. The second option, though, is forbidden in virtue of
the so-called {\it Earnshaw's theorem}\cite{Earnshaw,Wing-84} (for a
historical account on Samuel Earnshaw, the discoverer of this
theorem, we refer the interested reader to the work by William Scott
in this same journal.\cite{scott:AJP:1959}) According to this
theorem, in the absence of currents the modulus of the magnetic
field, $B \equiv |{\bf B}|$, has no local maxima
---the same result, as formerly shown by Earnshaw,\cite{Earnshaw}
applies to the electric field whenever the region under study is
free of charge.\cite{Jefimenko} Therefore, in practice the design of
magnetic traps reduces to finding configurations with a local
minimum, which are suitable to trap low-field seeker states.

%%%%%%%%%%%%%%%%%%%%%%%%%%%%%%%%%%%%%%%%%%%%%%%%%%%%%%%%%%%%%%%%%%%%%%%
%%%%%%%%%%%%%%%%%%%%%%%%%%%%%%%%%%%%%%%%%%%%%%%%%%%%%%%%%%%%%%%%%%%%%%%

\section{Types of magnetic traps}
\label{sec3}

From the discussion presented in Sec.~\ref{sec2}, it follows that
the basic ingredient involved in the design of an optimal trap
consists in finding an arrangement of coils and wires that produces
a magnetic field with the desired minimum. The purpose of this trap
is to confine an atomic or molecular cloud of a size much smaller
than the typical dimensions of the experimental magnetic
arrangements. Notice that typical cloud densities are $n \sim
10^{12}$--$10^{14}$~cm$^{-3}$, so that cloud sizes are in the range
$n^{-1/3} \sim 0.1$--$1$~$\mu$m, while the radii of coils and the
distances between coils and wires are of the order of centimeters!
Accordingly, one only needs to be concerned with the characteristics
of the magnetic field in a vicinity of its minimum, which allows for
simplifying series expansions around it, as seen below.

Typical magnetic traps display azimuthal symmetry, since the coils
playing the role of magnetic mirrors are of the same dimensions,
parallel one another, and centered around the azimuthal axis ($z$),
as seen in Fig.~\ref{fig2}. Accordingly, the minimum not only will
appear along the azimuthal axis, but at the origin of the coordinate
system. Of course, gravity will affect the position of this minimum,
by moving it downwards. However, in the analysis presented in this
work we will not include gravity for simplicity and also because it
gives rise to other types of effects, such as chaotic
dynamics.\cite{Metcalf} In order to obtain and analyze the
magnetic field around the local minimum, we can derive the exact
expression for the off-axis magnetic field or from the corresponding
magnetic scalar field, $\Phi$.
Then, a Taylor series expansion around the minimum is performed.
These two fields are related through the expression\cite{Panofsky}
\begin{equation}
 {\bf B} = - \nabla \Phi ,
 \label{eq5}
\end{equation}
since no currents are present between the coils.

The magnetic scalar potential $\Phi$ satisfies the Laplace equation
$\nabla^2 \Phi = 0$ (equivalent to $\nabla \cdot {\bf B} = 0$). Hence
it admits a general solution in spherical
coordinates,\cite{Jackson,Smythe,Panofsky,Jefimenko}
\begin{equation}
 \Phi = \sum_{\ell,m}
  \left[ C_{\ell,m} r^\ell + D_{\ell,m} r^{-(\ell+1)} \right]
  Y_\ell^m (\theta,\phi) ,
 \label{eq7}
\end{equation}
where $Y_\ell^m$ is the spherical harmonic function of degree $\ell$
and order $m$; the $C_{\ell,m}$ and $D_{\ell,m}$ coefficients
describe, respectively, the near and far field from the current
distribution, being dependent on its geometry and intensity.
Because trapping is assumed to occur in the near field and the system
displays azimuthal symmetry (no dependence on the azimuthal angle
$\phi$), Eq.~(\ref{eq7}) can be recast as
\begin{equation}
 \Phi = \sum_\ell C_\ell r^\ell P_\ell (\cos \theta ) ,
 \label{eq8}
\end{equation}
with $\ell$ labeling the multipole moment.

Now we already have the
essential tools to conduct our analysis of the different types of
magnetic traps.

%\begin{figure}
% \includegraphics[width=8.6cm]{J_PerezRios_Fig02.eps}
% \caption{\label{fig2}
%  Magnetic field configurations for a quadrupole trap (a) and a
%  Ioffe-Pritchard trap (b). The yellow arrows indicate the direction of
%  the currents flowing around the coils (blue torii) and through the
%  wires [gray bars along the $z$-direction in (b)].}
%\end{figure}

%%%%%%%%%%%%%%%%%%%%%%%%%%%%%%%%%%%%%%%%%%%%%%%%%%%%%%%%%%%%%%%%%%%%%%%

\subsection{Quadrupole trap}

The quadrupole trap is the simplest magnetic trap that can be
devised. It is obtained by positioning the two coils in
anti-Helmholtz configuration, i.e., with their currents flowing in
opposite directions [see Fig.~\ref{fig2}(a)].
The total field can be obtained as the
sum of the fields generated by each coil. So, in order to determine
the trapping field, let us first consider a current $I_c$ flowing along
a single coil of radius $a$.
The general expression for the magnetic field generated by such a
coil is analytical, expressible in terms of the complete elliptic
integrals.\cite{Jackson} As mentioned
above, the minimum will appear along the azimuthal axis, so we will
only be concerned with the behavior of the field along this axis. In
such a case, the magnetic scalar potential reads as
\begin{equation}
 \label{eq-9}
 \Phi_c(z) = - \frac{\mu_0 I_c}{2} \frac{z}{\sqrt{z^2 + a^2}} ,
\end{equation}
where $\mu_0$ is the vacuum magnetic permittivity. If now we take into
account the contribution coming from each coil, placed at $z=b$ and
$z=-b$, the total magnetic scalar potential is given by
\begin{equation}
 \label{eq-10}
  \Phi(z) = - \frac{\mu_0 I_c}{2}
   \left[ \frac{z-b}{\sqrt{(z-b)^2 + a^2}}
   - \frac{z+b}{\sqrt{(z+b)^2 + a^2}} \right] .
\end{equation}
As said above, the typical sizes for trapped clouds are much smaller
than the length scales associated with the magnetic arrangement.
Thus, under the assumption that $z \ll a, b$, Eq.~(\ref{eq-10}) can
be approximated by
\begin{equation}
 \label{eq-11}
 \Phi(z) = \frac{\mu_0 I_c}{2}
  \left[\frac{2b}{\left(a^2+b^2\right)^{1/2}}-\frac{3a^2bz^{2}}{
  \left(a^2+b^2\right)^{5/2}}\right] ,
\end{equation}
where we notice that only even powers of $z$ play a role.

After comparing Eq.~(\ref{eq8}) with (\ref{eq-11}), we can reasonably
argue that the sum in (\ref{eq8}) should also run only over even values
of $l$.
Indeed, we can also assume that the off-axis magnetic field is the
straightforward generalization of the on-axis one, given by
(\ref{eq-11}).
Under these hypotheses, we obtain $C_0 = \mu_0 I_c b/\sqrt{a^2+b^2}$,
$C_2 = - 3\mu_0 I_c a^2b/2(a^2+b^2)^{5/2}$, and $C_\ell = 0$ for
$\ell > 2$, at least at the order of the approximations here
considered.
Rewriting the resulting expression in cylindrical coordinates
($r=\sqrt{\rho^2+z^2}$ and $\cos \theta = z/r$), we find that
\begin{equation}
 \label{eq-13}
  \Phi(\rho,z) = C_0 + \frac{C_2}{2}\left(2z^2-\rho^2\right) .
\end{equation}
From Eq.~(\ref{eq5}), the magnetic field is readily found, which in
Cartesian coordinates reads as
\begin{equation}
 \label{eq-14}
  {\bf B} = - C_2 \left( x {\bf u}_{x}
   + y {\bf u}_{y} - 2z {\bf u}_{z} \right) .
\end{equation}
As seen, the gradient of this field with vanishing divergence is ruled
by $C_2$, i.e., the size and distance between the coils. As it can be
seen from the modulus of the magnetic field,
\begin{equation}
 B = \| {\bf B} \| = C_2 \sqrt{x^2 + y^2 + 4z^2} ,
 \label{eq-14b}
\end{equation}
this trap can never be perfectly harmonic.

According to Eq.~(\ref{eq-14b}), the magnetic field of a qua\-dru\-po\-le
trap has a zero minimum at the origin. Hence, if the trapped particles
reach the minimum, the energy spacing between Zeeman levels becomes
rather reduced (see Fig.~\ref{fig1}). This increases the probability
of non-adiabatic transitions among these energy levels, leading to a
loss of particles in the trap. This scape mechanism is known as {\it
spin-flip Majorana transition}.\cite{Majorana,brink}

%%%%%%%%%%%%%%%%%%%%%%%%%%%%%%%%%%%%%%%%%%%%%%%%%%%%%%%%%%%%%%%%%%%%%%%

\subsection{TOP trap}

The TOP trap is a modification of the quadrupole trap, which skips
the inconvenience of the zero-field minimum. More specifically, this
modification consists of adding a time-dependent rotating bias field
in the $XY$-plane, so that
%
%\begin{eqnarray}
% {\bf B} & = & (C_2 x + B_b \cos \omega t){\bf u}_{x}
%  \nonumber \\
%  & & + ( C_2 y + B_b \sin \omega t){\bf u}_{y} - 2 C_2 z{\bf u}_{z} .
% \label{time-field}
%\end{eqnarray}
\be
 {\bf B} = (C_2 x + B_b \cos \omega t){\bf u}_{x}
 + ( C_2 y + B_b \sin \omega t){\bf u}_{y} - 2 C_2 z{\bf u}_{z} .
 \label{time-field}
\ee
This makes the minimum to revolve around the center of the trap in a
circle of radius $R = B_b/C_2$.
This is better seen by computing the modulus of this magnetic field,
\be
 B = \sqrt{(C_2 x + B_b \cos \omega t)^2
   + (C_2 y + B_b \sin \omega t)^2 + 4 C_2^2 z^2} .
 \label{time-field-modulus}
\ee
The bias frequency, $\omega$, provides us with information about the
working regime. In order to select an optimal trapping regime,
$\omega$ has to satisfy two conditions:
\begin{enumerate}
 \item It should be larger than the oscillation frequency of the
 trapped particles ($\sim 100$~Hz), so that they will feel an effective
 time-averaged magnetic field.

 \item It should be smaller than the frequency associated with
 the transition between two consecutive internal quantum states
 ($\sim 10^6$~Hz) to prevent particle losses by spin-flip Majorana
 transitions.
\end{enumerate}

In the particular case that we are interested in, we have
$r = \sqrt{x^2 + y^2 + z^2} \ll R$, and Eq.~(\ref{time-field})
can be approximated by
%
%\begin{eqnarray}
% B = B_b & + & \frac{C_2^2}{2 B_b^2} \ \! (x^2 + y^2 + 4z^2)
%   \nonumber \\
% & + & \frac{C_2}{B_b} \ \! (x \cos \omega t + y \sin \omega t)
%   \nonumber \\
% & - & \frac{C_2^2}{2 B_b^2}
%   \ \! (x \cos \omega t + y \sin \omega t)^2 .
%\end{eqnarray}
\be
 B = B_b + \frac{C_2^2}{2 B_b^2} \ \! (x^2 + y^2 + 4z^2)
 + \frac{C_2}{B_b} \ \! (x \cos \omega t + y \sin \omega t)
 - \frac{C_2^2}{2 B_b^2} \ \! (x \cos \omega t + y \sin \omega t)^2 .
\ee
From this expression, we can now obtain the effective time-averaged
magnetic field felt by the trapped particles,
\begin{equation}
 \bar{B} = \frac{\omega}{2\pi} \int_0^{2\pi/\omega} B(t) dt
  = B_b + \frac{C_2^2}{4B_b} \ \! (x^2 + y^2 + 8z^2) .
 \label{TOP}
\end{equation}
As it can be noticed, the effect of the bias is to generate an
(effective) anisotropic harmonic potential, with an energy origin
shifted upwards in energy ($B_b$) and frequencies $\omega_x =
\omega_y = \sqrt{-\mu C_2^2/2mB_b}$ and $\omega_z = \sqrt{- 4\mu
C_2^2/mB_b}$, where $m$ is the mass of the trapped species. These
frequencies arise after substitution of Eq.~(\ref{TOP}) in the
expression for the magnetic interaction energy (\ref{eq-1}). We thus
obtain an effective interaction potential, such that $\mu = - dE/dB
< 0$ for low-field seeker states according to Fig.~\ref{fig1}.

%%%%%%%%%%%%%%%%%%%%%%%%%%%%%%%%%%%%%%%%%%%%%%%%%%%%%%%%%%%%%%%%%%%%%%%

\subsection{Ioffe-Pritchard trap}

In the traps that we have just analyzed, we were mainly concerned
about trapping conditions along the azimuthal direction.
We also ended up getting some trapping along the radial direction,
although this was not the main goal.
In order to determine
specific trapping conditions along this direction, it is
necessary to introduce an additional constraining field. In
practice, this is achieved by incorporating a series of wires around
the two coils, as it happens in Ioffe-Pritchard traps
[see Fig.~\ref{fig2}(b)]. In this trap, four wires are suited at the
corners of a square, with the currents flowing along adjacent wires
being of opposite sign (see Fig.~\ref{fig3}).

Let us start again by first considering the case of a current $I_w$
flowing along a single, straight wire. The magnetic field generated
by this wire is given by
\begin{equation}
 \label{IP-1}
 {\bf B}=- \frac{\mu_0 I_w}{2\pi}\left(
 \frac{y{\bf u}_x - x{\bf u}_y}{x^2 + y^2} \right) .
\end{equation}
This expression can be easily obtained from Ampere's law (in this
case, determining the magnetic field from this law is simpler than
doing it through the magnetic scalar potential) and taking into
account the transformation relations of unitary vectors from
cylindrical to Cartesian coordinates
\be
 \begin{array}{rcl}
 {\bf u}_\rho & = & \cos \phi {\bf u}_x + \sin \phi {\bf u}_y \\
 {\bf u}_\phi & = & -\sin \phi {\bf u}_x + \cos \phi {\bf u}_y
 \end{array} .
\ee
(Actually, from symmetry considerations, we can see that only ${\bf
u}_\phi$ is necessary.)

%\begin{figure}
% \begin{center}
% \includegraphics[width=8.6cm]{J_PerezRios_Fig03.eps}
% \caption{\label{fig3}
%  Arrow map of the magnetic field (\ref{b-wires}), generated by the
%  four parallel wires of Fig.~\ref{fig2}(b).
%  The length of the arrows gives the intensity of field, while the
%  direction of the currents is indicated by a dot (outwards flow) or
%  a cross (inwards flow).}
% \end{center}
%\end{figure}

The total magnetic field generated by the four wires can be now
readily obtained from (\ref{IP-1}) if we evaluate this expression at
the points $(x+d,y+d)$, $(x-d,y+d)$, $(x-d,y-d)$, and $(x+d,y-d)$,
instead of $(x,y)$, and we add the four resulting fields. Then, as
before, assuming $d \gg x, y$, we obtain
\begin{equation}
 {\bf B} = \frac{\mu_0 I_w}{\pi d^2}
  \left( x {\bf u}_x - y {\bf u}_y \right) .
 \label{b-wires}
\end{equation}
A cross-section of this field is displayed in Fig.~\ref{fig3}, where
the length of the (blue) arrows is proportional to the field
intensity. As it can be seen, we observe that these arrows point
towards the origin (upwards or downwards) along the axis $x=0$, and
outwards the origin (leftwards or rightwards) along the axis $y=0$.
From a dynamical point of view,\cite{jordan} the first axis
is called {\it stable} and the second, {\it unstable}; the origin is
called a {\it saddle} or {\it hyperbolic point}. To better
understand the dynamical implications of this fact, consider a
system directly acted by the magnetic field (\ref{b-wires}). If this
system is released at some point $y_0$ and $x_0=\epsilon$, with
$\epsilon \to 0$, it will fall towards $y=0$, separating from the
$x=0$ axis slowly initially and then very rapidly as it approaches
the unstable axis, $y=0$. These points are, for example, at the
origin of chaotic behaviors or high sensitivity to initial conditions
(e.g., exponential divergence between two trajectories that are
launched from very close initial conditions).

The presence of the saddle point is a direct consequence of the
quadrupolar nature of the arrangement in the radial direction.
However, the same point becomes a stable minimum for the modulus of
the magnetic field,
\begin{equation}
 B = \frac{\mu_0 I_w}{\pi d^2}\ \! \sqrt{ x^2 + y^2} .
 \label{bmod-wires}
\end{equation}
In this case, again from a dynamical point of view, this means that
a particle acted by (\ref{bmod-wires}) instead of ${\bf B}$ will
always fall down to the origin, like an oscillator. This minimum can
be seen in Fig.~\ref{fig4}, where the magnetic field
(\ref{bmod-wires}) is plotted along the $x$-axis [i.e., the unstable
direction for (\ref{b-wires})].

In the field of plasma confinement, this particular type of magnetic
configuration, formed by the four bars, is known as Ioffe
bars.\cite{Gott-62} In analogy to quadrupole traps, spin-flip
Majorana transitions can also occur here due to the presence of a
vanishing field minimum. To overcome this inconvenience, a
homogeneous magnetic field along the azimuthal direction, $B_h{\bf
u}_z$ is introduced, so that
\begin{equation}
 {\bf B} = \frac{\mu_0 I_w}{\pi d^2}
  \left( x {\bf u}_x - y {\bf u}_y + z_0 {\bf u}_z \right) ,
 \label{bh-wires}
\end{equation}
with $B_h = \mu_0 I_w z_0/\pi d^2$. The modulus of this field is
\begin{equation}
 B = \frac{\mu_0 I_w}{\pi d^2}\ \! \sqrt{ x^2 + y^2 + z_0^2} ,
 \label{bmodh-wires}
\end{equation}
which has a non-vanishing minimum due to the shift induced by the
above homogeneous field (see Fig.~\ref{fig4}), this avoiding the
losses associated with spin-flip Majorana transitions. Obviously,
this configuration changes the behavior of the magnetic field close
to the minimum, from a linear dependence to a quadratic.

%\begin{figure}
%\begin{center}
%\includegraphics[width=8.6cm]{J_PerezRios_Fig04.eps}
% \caption{\label{fig4}
%  Modulus of the magnetic field described by Eq.~(\ref{bmodh-wires}).
%  This is the field generated by the four parallel wires of a
%  Ioffe-Pritchard trap [see Fig.~\ref{fig2}(b)] along the $y=0$
%  direction for three different values of the homogeneous field:
%  $B_h = 0$~mT (blue dotted line), $B_h = 25$~mT (red dashed line),
%  and $B_h = 50$~mT (green solid line).
%  The parameters considered are typical values: $I_w=10^4$~A and
%  $d=0.01$~m.
%  The horizontal line in each case indicates the limit where the
%  magnetic field is 10~mT above the corresponding minimum.}
%\end{center}
%\end{figure}

The homogeneous magnetic field as well as the confinement along the
azimuthal direction can be generated if the two coils are arranged in
Helmholtz configuration, i.e., making the intensities in each coil to
flow in the same direction [see Fig.~\ref{fig2}(b)]. In this case, the
on-axis magnetic scalar potential reads as
\begin{equation}
 \label{IP-2}
 \Phi(z) = - \frac{\mu_0 I_c}{2} \left[
  \frac{z-b}{\sqrt{(z-b)^2 + a^2}} + \frac{z+b}{\sqrt{(z+b)^2 + a^2}}
  \right] .
\end{equation}
Then, assuming again $z \ll a,b$ and expanding for small $z$, we
find
\begin{equation}
 \label{IP-3}
 \Phi(z) = - \frac{\mu_0 I_c}{2} \left[
  \frac{2a^2z}{\left(a^2+b^2\right)^{3/2}}
  - \frac{(a^4-4a^2b^2)z^{3}}{\left(a^2+b^2\right)^{7/2}}\right] ,
\end{equation}
which depends on odd of powers of $z$ due to the Helmholtz
configuration of the coils. When comparing with Eq.~(\ref{eq8}), we
find that now the sum will run over odd values of $\ell$. More
specifically, we will have $C_1 = -\mu_0 I_c a^2/(a^2 + b^2)^{3/2}$,
$C_3 = \mu_0 I_c a^2 (a^2 - 4b^2)/2(a^2+b^2)^{7/2}$, and $C_\ell = 0$
for $\ell > 3$ (at the order of approximation considered). By using
cylindrical coordinates, Eq.~(\ref{IP-3}) can be recast as
\begin{equation}
 \label{IP-4}
 \Phi(\rho,z) = C_1 z + \frac{C_3}{2} \left( 2z^3 - 3z\rho^2 \right) .
\end{equation}
The corresponding magnetic field in Cartesian coordinates reads as
\begin{equation}
% {\bf B} = - C_1 {\bf u}_z + 3C_3 z \left( x {\bf u}_x
%  + y {\bf u}_y - z {\bf u}_z \right)
%  + \frac{3}{2}\ \! C_3 \left( x^2 + y^2 \right) {\bf u}_z .
 {\bf B} = 3C_3 z \left( x{\bf u}_x + y {\bf u}_y \right)
  - \left[ C_1
  - \frac{3}{2}\ \! C_3 \left( x^2 + y^2 - 2z^2 \right) \right]
  {\bf u}_z .
 \label{b-IP}
\end{equation}

In Eq.~(\ref{b-IP}) we notice confinement along the azimuthal
direction. This becomes more apparent if we consider the field along
the azimuthal axis, for $x=0$ and $y=0$,
\begin{equation}
 \label{bz-IP}
 B_z = -C_1 - 3C_3 z^2
\end{equation}
(in this case, $B_x=B_y=0$). The first term is a homogeneous field,
while the second is a quadratic trapping term related to the
curvature of the magnetic field. Contrary to $C_2$ in the quadrupole
trap, here $C_3$ has a sign that depends on the arrangement
geometry, i.e., the radius of the coils, $a$, and their distance,
$2b$. Thus, by selecting the coil parameters $a$ and $b$ in a
convenient manner, one can generate more tightly trapping
potentials, and also change their curvature (or even suppress it).
In this sense, one could benefit from the relation $C_3 \propto a -
2b$: by choosing $a = 2b$, we get the homogeneous field along the
azimuthal axis needed to avoid spin-flip Majorana transitions in the
radial direction, with $B_h = -C_1 = 8\mu_0 I_c/5\sqrt{5}a$.

Now, given a particular arrangement of currents, a maximum magnetic
field $B_{\rm max}$ is obtained. According to Eq.~(\ref{eq-1}), this
value determines the largest amount of energy accessible to the
particles inside the trap, $E_{\rm max} = \mu (B_{\rm max} - B_{\rm
min})$. Any particle with kinetic energy larger than $E_{\rm max}$
will escape from the trap. In this sense, $E_{\rm max}$ is defined
as the trap depth. It is one of the parameters that has to be taken
into account when designing magnetic traps, since it will give us an
estimate of the number of trapped particles from among an ensemble
characterized by a certain velocity
distribution.\cite{Hess-86,Ketterle-96} In Fig.~\ref{fig4}, for
example, this maximum amount of energy (horizontal lines) is
expressed in terms of the corresponding field intensity. In
particular, we have considered $B_{\rm max} - B_{\rm min} = 10$~mT.

%%%%%%%%%%%%%%%%%%%%%%%%%%%%%%%%%%%%%%%%%%%%%%%%%%%%%%%%%%%%%%%%%%%%%%%
%%%%%%%%%%%%%%%%%%%%%%%%%%%%%%%%%%%%%%%%%%%%%%%%%%%%%%%%%%%%%%%%%%%%%%%

\section{Discussion}
\label{sec4}

Once we have seen the essential elements involved in the design of
the three kind of traps used to produce and confine BECs, it is
interesting to establish a comparative analysis in terms of their
confinement efficiency.
In this regard, we would like to specify that, within this context,
by efficiency we mean the capability of a particular magnetic
configuration to confine the particle cloud within a certain space
region.
This analysis can be carried out by investigating the magnetic field
around the trap minimum.

As mentioned above, spin-flip Majorana transitions constitute a loss
mechanism that has to be taken into account in order to set some
control on the efficiency of a magnetic trap. The rate of
non-adiabatic transitions between trapped and non trapped states
increases when the magnetic configuration has a zero-field minimum,
as happens in the quadrupole trap.
Therefore, for low field intensities what one expects is that
particle losses per time unit will be higher in quadrupole traps than
in TOP or Ioffe-Pritchard ones.

Let us consider, though, the behavior of the field close to the (trap)
minimum and analyze the energy of the particles inside the trap.
Following Eq.~(\ref{eq-1}), a suitable
estimate of this energy is given by\cite{Petrich-95}
\begin{equation}
 E \sim \mu B_c \left( \frac{n^{-1/3}}{a} \right)^{\ell-1} ,
\end{equation}
where $\mu$ is the modulus of the atomic magnetic moment and $B_c$ is
the magnetic field generated by the coils. As we saw in Sec.~\ref{sec2},
the maximum multipole moment for quadrupole traps is $\ell = 2$, while
for TOP and Ioffe-Pritchard traps is $\ell = 3$.
Accordingly, the interaction potential for the former (with vanishing
minimum) is a factor $n^{-1/3}/a$ larger than for the latter (with
non-vanishing minima).
Typically, this factor ranges between 10 and 1,000, making the
quadrupole trap to be more efficient than the TOP and Ioffe-Pritchard
ones (in the sense indicated above).

%\begin{figure}
%\begin{center}
%\includegraphics[width=8.6cm]{J_PerezRios_Fig05.eps}
% \caption{\label{fig5}
%  (a) Modulus of the magnetic field along the $x$-direction (for
%  $y=z=0$) for a quadrupole trap (black dotted line), a TOP
%  trap (red dashed line), and a Ioffe-Pritchard trap (blue solid
%  line).
%  (b) Same as in panel (a), but along the $z$-direction (for $x=y=0$).
%  (c) Effective particle density associated with these traps as a
%  function of the temperature [the same convention of colors and line
%  types as in parts (a) and (b) has been followed].
%  This density has been obtained as the inverse of the trap effective
%  volume obtained from Eq.~(\ref{effvol}).
%  The numerical parameters considered in the calculations of the graphs
%  displayed in the three panels are:
%  Quadrupole trap: $a = 0.01$~m, $b = 0.0125$~m, and $I_c = 3,000$~A;
%  TOP trap: $a = 0.01$~m, $b = 0.0125$~m, $I_c = 3,000$~A, and $B_b = 15$~mT;
%  Ioffe-Pritchard trap: $a = 0.01$~m, $b = 0.0125$~m, $d = 0.01$~m,
%  $I_c=300$~A, and $I_w = 10^4$~A.}
%\end{center}
%\end{figure}

We can also estimate the cloud density inside the magnetic trap, which
can be expressed as a Maxwell-Botzmann distribution,
\begin{equation}
 n({\bf r}) = n_0 e^{-U({\bf r})/k_B T} ,
\end{equation}
where $n_0$ is the density at the center of the trap and $U({\bf
r})$ is the trapping potential.
The latter is obtained from the magnetic energy difference between a
given value of the field and its minimum, i.e., from Eq.~(\ref{eq-1}),
$U({\bf r}) = \Delta E = \mu [B({\bf r}) - B_{\rm min}]$, with $\mu$
denoting the modulus of the magnetic dipole moment.
The total number of trapped particles is thus given by
\begin{equation}
 N = n_0 \int e^{-U({\bf r})/k_B T} d{\bf r} ,
 \label{effvol}
\end{equation}
where the integral can be defined as the effective volume occupied by
the particle cloud, $V_{\rm eff} \equiv \int e^{-U({\bf r})/k_B T}
d{\bf r}$.
In Fig.~\ref{fig5}(a) we show the profile of the magnetic field along
the $x$-direction (with $y=z=0$) for the three traps analyzed in the
previous section (the parameters involved in the calculations are given
in the figure caption); similarly, in Fig.~\ref{fig5}(b) the same is
done along the $z$-direction (with $x=y=0$).
As we are going to see, independently of the minimum value, in this
case the particle density is governed by the trap topology (curvature).
Thus, for the quadrupole trap, from (\ref{eq-14b}) the trapping
potential reads as
\begin{equation}
 U({\bf r}) = \mu C_2 \sqrt{x^2 + y^2 + 4z^2}
\end{equation}
and therefore $V_{\rm eff} \sim T^3$.
However, if the behavior of the magnetic field is smoother around the
trap minimum, as in TOP and Ioffe-Pritchard traps, the dependence on
temperature will change.
Taking as an example the TOP trap, with trapping potential
\begin{equation}
 U({\bf r}) = \frac{\mu C_2^2}{4B_b} \ \! (x^2 + y^2 + 8z^2) ,
\end{equation}
we find $V_{\rm eff} \sim T^{3/2}$.

From a practical point of view, we know that in order to reach the BEC
state, the phase-space density, which goes as $n\lambda_{\rm dB}^3$,
has to be increased by decreasing the particle temperature. Now, given
a constant number of trapped particles, since the particle density
$n_0$ is inversely proportional to $V_{\rm eff}$, from the above two
behaviors we find that effectively it will increase as $T$
decreases.
The variation of the density as a function of the temperature is
displayed in Fig.~\ref{fig5}(c) for the three types of magnetic traps
considered in this work.
For the chosen parameters (see figure caption for details), we notice
that, at high temperatures, the particle density in a Ioffe-Pritchard
trap is higher than in the other two ones, while as temperature
decreases the opposite behavior is observed.
A crossover for these trends is observed at $T \approx 3$~mK (other
crossovers can also be seen at other temperatures).

Such a behavior with temperature is related to the shape of the
magnetic field around the minimum, which has an important influence on
the confinement or compression of the swarm of trapped particles per
volume unit.
More specifically, this compression is connected to the particle
kinetic energy through the temperature of the sample.
That is, at higher temperatures particles are able to visit regions with
higher values of the potential energy, while at lower temperatures only
regions closer to the minimum can be explored.
This behaviors can be easily inferred by inspecting Fig.~\ref{fig5}(a)
in combination with Fig.~\ref{fig5}(b).
Thus, for low temperatures, in spite of the high confinement displayed
by the Ioffe-Pritchard trap along the $x$-direction [see
Fig.~\ref{fig5}(a)], along the $z$-direction (for the $y$-direction it
would be the same as for the $x$ one) the confinement is very poor
[see Fig.~\ref{fig5}(b)].
On the contrary, for the quadrupole trap the confinement is optimal
along either direction, thence the confining effect is higher in this
type of trap than in the other two, as seen in Fig.~\ref{fig5}(c).
Now, as one increases temperature, this behavior changes, because the
curvature of the Ioffe-Pritchard trap starts playing a role.
As for TOP traps, they show a similar behavior to Ioffe-Pritchard
traps, but with a poorer confining performance [about one order of
magnitude up to $T \approx 3$~mK, approximately; see
Fig.~\ref{fig5}(c)], except at relatively high temperatures.

From the above analysis on Fig.~\ref{fig5}, therefore, one can now get
a simple idea about why Ioffe-Pritchard traps are more commonly used
than the other two types.
As seen, they provide a relatively high confinement at low
temperatures, removing at the same time the inconvenience of the
zero-field minimum that leads to spin-flip transitions.
This is, indeed, in accordance with current design considerations of
neutral-particle traps.\cite{butterworth}

%%%%%%%%%%%%%%%%%%%%%%%%%%%%%%%%%%%%%%%%%%%%%%%%%%%%%%%%%%%%%%%%%%%%%%%
%%%%%%%%%%%%%%%%%%%%%%%%%%%%%%%%%%%%%%%%%%%%%%%%%%%%%%%%%%%%%%%%%%%%%%%

\section{Concluding remarks}
\label{sec5}

Given the relevance of magnetic trapping in ultracold physics, here
we have presented a comprehensive analysis and discussion of the
basic physics and physical properties related to three of the most
commonly used magnetic traps considered in Bose-Einstein
condensation: the quadrupole trap, the time-averaged orbiting
potential trap, and the Ioffe-Pritchard trap. It has been shown that
from relatively simple considerations about magnetic field generated
by the different elements that constitute the trap, one can
determine the trapping conditions and efficiency of these devices.

The considerable reduction of technicalities in these derivations
presents also a pedagogic advantage, for they can result of interest
in elementary courses on classical electromagnetism, even knowing
nothing about the intrinsic quantum nature of the trapped particles,
namely the Bose-Einstein condensates.
Actually, the type of estimates shown here can be combined with further
classical and quantum-mechanical dynamical studies,\cite{gov} giving
rise to a rather balanced combination of elementary concepts, easy to
handle by students and, in general, anyone interested in this timely
research field.

%%%%%%%%%%%%%%%%%%%%%%%%%%%%%%%%%%%%%%%%%%%%%%%%%%%%%%%%%%%%%%%%%%%%%%%

\acknowledgments

The authors would like to thank two anonymous referees for their
valuable remarks.
Partial support from the IFRAF (France) and the Ministerio de
Econom\'{\i}a y Competitividad (Spain) under Projects FIS2010-22082
and FIS2011-29596-C02-01 is acknowledged. A. S. Sanz would also like
to thank the Ministerio de Econom\'{\i}a y Competitividad for a
``Ram\'on y Cajal'' Research Grant and the University College London
for its kind hospitality during the elaboration of this work.

%%%%%%%%%%%%%%%%%%%%%%%%%%%%%%%%%%%%%%%%%%%%%%%%%%%%%%%%%%%%%%%%%%%%%%%
%%%%%%%%%%%%%%%%%%%%%%%%%%%%%%%%%%%%%%%%%%%%%%%%%%%%%%%%%%%%%%%%%%%%%%%

%\begin{thebibliography}{99}
%\eprint{}

%

%\end{document}

%%%%%%%%%%%%%%%%%%%%%%%%%%%%%%%%%%%%%%%%%%%%%%%%%%%%%%%%%%%%%%%%%%%%%%%
%%%%%%%%%%%%%%%%%%%%%%%%%%%%%%%%%%%%%%%%%%%%%%%%%%%%%%%%%%%%%%%%%%%%%%%

\newpage % Caption of figures

\section*{List of figure captions}

\begin{figure}[!h]
%\centering
% \includegraphics[width=8cm]{J_PerezRios_Fig01.eps}
 \caption{\label{fig1}
  Zeeman splitting of the $^{16}$O$_{2}(^{3}\Sigma_{g}^{-})$ molecular
  energy levels (in temperature units) as a function of the magnetic
  field intensity.
  High-field and low-field seeker states are denoted, respectively,
  by red solid lines and blue dashed lines; gray dotted lines
  illustrate the appearance of Zeeman splitting in higher energy
  levels.
  These results have been obtained using realistic values in the
  simulation:\cite{Mizushima}
  $B_e = 1.438$~cm$^{-1}$ (rotational constant), $\gamma_{\rm SR} =
  -0.0089$~cm$^{-1}$ (spin-rotation interaction), and $\gamma_{\rm SS}
  = 1.985$~cm$^{-1}$ (spin-spin coupling).}
\end{figure}

\begin{figure}[!h]
%\centering
% \includegraphics[width=8.5cm]{J_PerezRios_Fig02.eps}
 \caption{\label{fig2}
  Magnetic field configurations for a quadrupole trap (a) and a
  Ioffe-Pritchard trap (b). The yellow arrows indicate the direction of
  the currents flowing around the coils (blue torii) and through the
  wires [gray bars along the $z$-direction in (b)].}
\end{figure}

\begin{figure}[!h]
 \begin{center}
 \caption{\label{fig3}
  Arrow map of the magnetic field (\ref{b-wires}), generated by the
  four parallel wires of Fig.~\ref{fig2}(b).
  The length of the arrows gives the intensity of field, while the
  direction of the currents is indicated by a dot (outwards flow) or
  a cross (inwards flow).}
 \end{center}
\end{figure}

\begin{figure}[!h]
\begin{center}
 \caption{\label{fig4}
  Modulus of the magnetic field described by Eq.~(\ref{bmodh-wires}).
  This is the field generated by the four parallel wires of a
  Ioffe-Pritchard trap [see Fig.~\ref{fig2}(b)] along the $y=0$
  direction for three different values of the homogeneous field:
  $B_h = 0$~mT (blue dotted line), $B_h = 25$~mT (red dashed line),
  and $B_h = 50$~mT (green solid line).
  The parameters considered are typical values: $I_w=10^4$~A and
  $d=0.01$~m.
  The horizontal line in each case indicates the limit where the
  magnetic field is 10~mT above the corresponding minimum.}
\end{center}
\end{figure}

\begin{figure}[!h]
\begin{center}
 \caption{\label{fig5}
  (a) Modulus of the magnetic field along the $x$-direction (for
  $y=z=0$) for a quadrupole trap (black dotted line), a TOP
  trap (red dashed line), and a Ioffe-Pritchard trap (blue solid
  line).
  (b) Same as in panel (a), but along the $z$-direction (for $x=y=0$).
  (c) Effective particle density associated with these traps as a
  function of the temperature [the same convention of colors and line
  types as in parts (a) and (b) has been followed].
  This density has been obtained as the inverse of the trap effective
  volume obtained from Eq.~(\ref{effvol}).
  The numerical parameters considered in the calculations of the graphs
  displayed in the three panels are:
  Quadrupole trap: $a = 0.01$~m, $b = 0.0125$~m, and $I_c = 3,000$~A;
  TOP trap: $a = 0.01$~m, $b = 0.0125$~m, $I_c = 3,000$~A, and $B_b = 15$~mT;
  Ioffe-Pritchard trap: $a = 0.01$~m, $b = 0.0125$~m, $d = 0.01$~m,
  $I_c=300$~A, and $I_w = 10^4$~A.}
\end{center}
\end{figure}

%%%%%%%%%%%%%%%%%%%%%%%%%%%%%%%%%%%%%%%%%%%%%%%%%%%%%%%%%%%%%%%%%%%%%%%
%%%%%%%%%%%%%%%%%%%%%%%%%%%%%%%%%%%%%%%%%%%%%%%%%%%%%%%%%%%%%%%%%%%%%%%

\vspace{10cm}

\begin{figure}[!h]
 \begin{center}
 \end{center}
\end{figure}

\newpage

\section*{Figures}
\setcounter{figure}{0}

\vspace{2cm}

\begin{figure}[!h]
 \begin{center}
  \includegraphics[width=8.6cm]{J_PerezRios_Fig01.eps}
   \caption{}
 \end{center}
\end{figure}

\newpage

\begin{figure}[!h]
 \begin{center}
  \includegraphics[width=8.6cm]{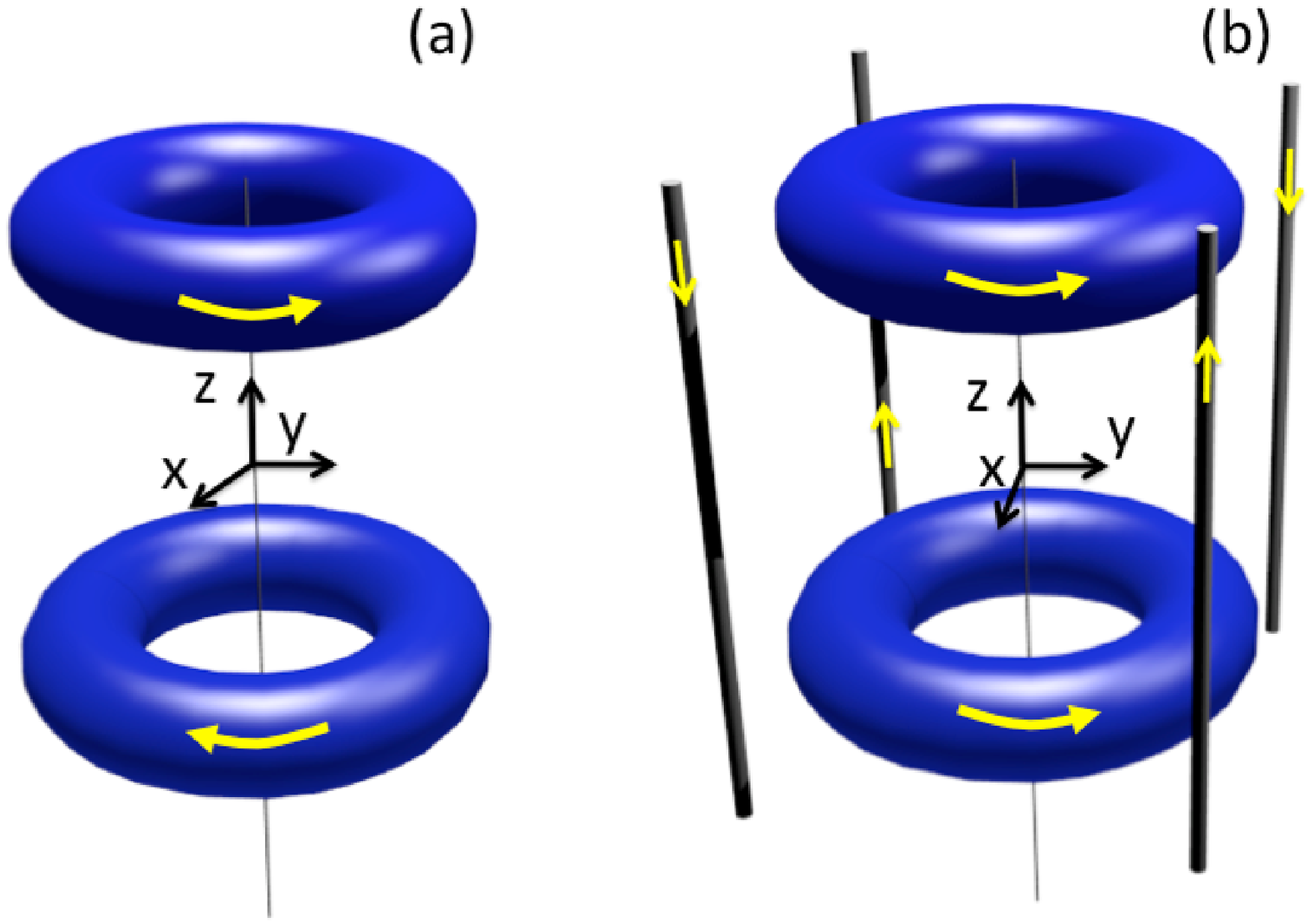}
   \caption{}
 \end{center}
\end{figure}

\newpage

\begin{figure}[!h]
 \begin{center}
  \includegraphics[width=8.6cm]{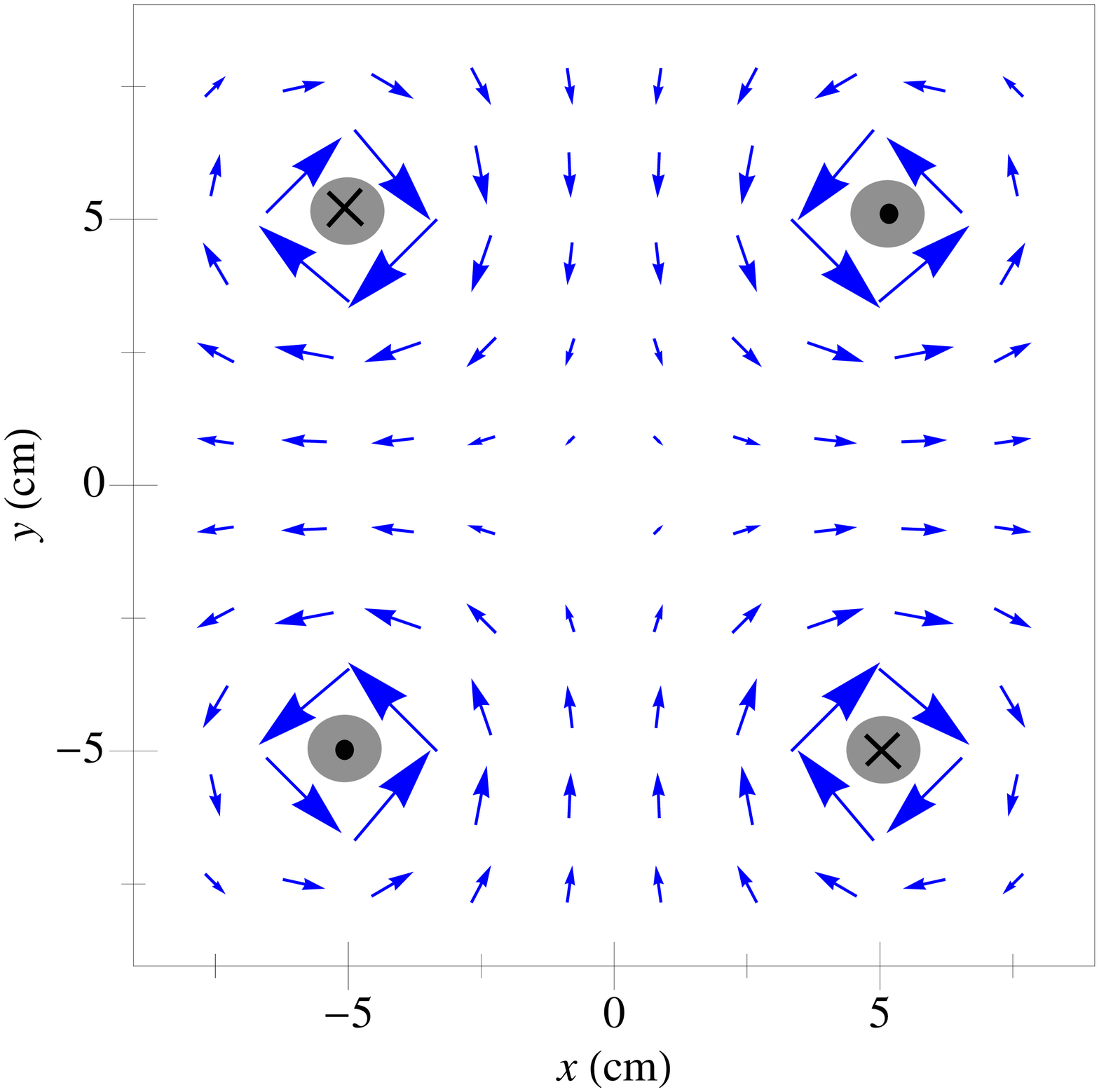}
   \caption{}
 \end{center}
\end{figure}

\newpage

\begin{figure}[!h]
 \begin{center}
  \includegraphics[width=8.6cm]{J_PerezRios_Fig04.eps}
   \caption{}
 \end{center}
\end{figure}

\newpage

\begin{figure}
 \begin{center}
  \includegraphics[width=8.6cm]{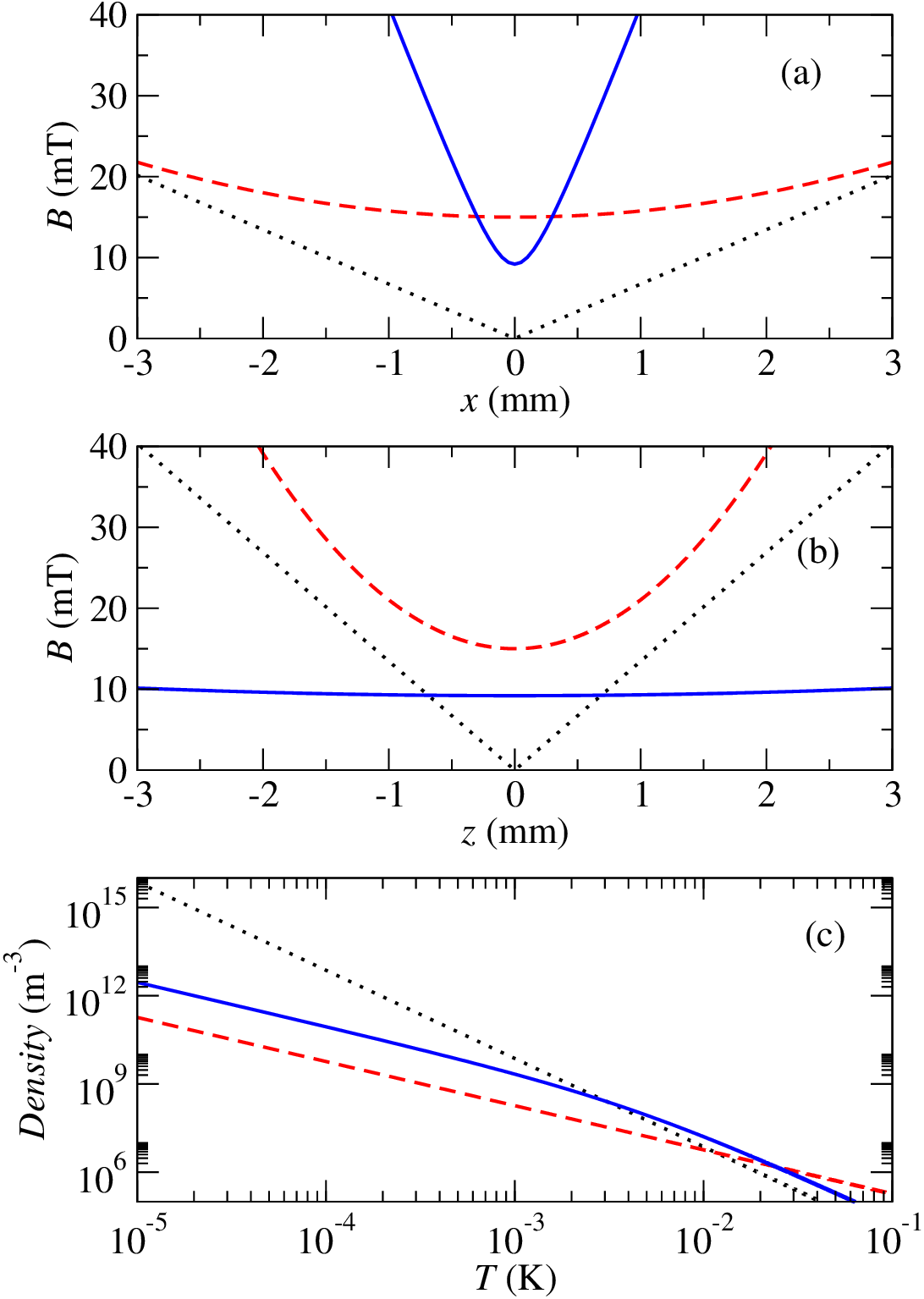}
   \caption{}
 \end{center}
\end{figure}

\end{document}